\documentclass[journal]{IEEEtran}
\usepackage{epsfig}
\usepackage{epstopdf}
 
\begin{document}

\title{Internet of Bio-Nano Things: A Review of Applications, Enabling Technologies and Key Challenges}

\author{Murat Kuscu,~\IEEEmembership{Member,~IEEE}
        and Bige D. Unluturk,~\IEEEmembership{Member,~IEEE}%% <-this % stops a space        
\thanks{
Murat Kuscu is with the Department of Electrical and Electronics Engineering, Koc University, Rumelifeneri Yolu 34450, Sariyer, Istanbul, Turkey (e-mail: mkuscu@ku.edu.tr). }
\thanks{Bige D. Unluturk is with the Department of Electrical \& Computer Engineering and Biomedical Engineering, 775 Woodlot Dr., East Lansing, 48823, MI, USA (e-mail: unluturk@msu.edu)}
\thanks{The work of Murat Kuscu was supported in part by the European Union’s Horizon 2020 Research and Innovation Programme through the Marie Skłodowska-Curie Individual Fellowship under Grant Agreement 101028935 and by The Scientific and Technological Research Council of Turkey (TUBITAK) under Grant \#120E301.}}% <-this % stops a space

\IEEEpeerreviewmaketitle
\maketitle

\begin{abstract}
Internet of Bio-Nano Things (IoBNT) is envisioned to be a heterogeneous network of nanoscale and biological devices, so called Bio-Nano Things (BNTs), communicating via non-conventional means, e.g., molecular communications (MC), in non-conventional environments, e.g., inside human body. The main objective of this emerging networking framework is to enable direct and seamless interaction with biological systems for accurate sensing and control of their dynamics in real time. This close interaction between bio and cyber domains with unprecedentedly high spatio-temporal resolution is expected to open up vast opportunities to devise novel applications, especially in healthcare area, such as intrabody continuous health monitoring. There are, however, substantial challenges to be overcome if the enormous potential of the IoBNT is to be realized. These range from developing feasible nanocommunication and energy harvesting techniques for BNTs to handling the big data generated by IoBNT. In this survey, we attempt to provide a comprehensive overview of the IoBNT framework along with its main components and applications. An investigation of key technological challenges is presented together with a detailed review of the state-of-the-art approaches and a discussion of future research directions.
\end{abstract}

\section{Introduction} 
\label{sec:intro}
As the Internet of Things (IoT) approaches technological maturity with growing number of applications on the market, new integrative ideas emerge to push the current boundaries of IoT and extend its application range. One such approach follows a holistic view and regards the universe as an interconnected entity which is to be observed, understood, and manipulated with new information and communication technologies (ICT). At the center of this approach lies an emerging ICT framework, the Internet of Bio-Nano Things (IoBNT), envisioning the heterogeneous collaborative networks of natural and artificial nano-biological functional devices (e.g., engineered bacteria, human cells, nanobiosensors), seamlessly integrated to the Internet infrastructure  \cite{akyildiz2015internet}. IoBNT is positioned to extend our connectivity and control over non-conventional domains (e.g., human body) with unprecedented spatio-temporal resolution, enabling paradigm-shifting applications, particularly in the healthcare domain, such as intrabody continuous health monitoring and theranostic systems with single molecular precision. 

The broad application prospects of IoBNT have attracted significant research interest at the intersection of ICT, bio-nanotechnology, and medical sciences, with the great majority of studies directed towards (i) the design and implementation of Bio-Nano Things (BNTs) \cite{kuscu2018survey, bi2021survey}, (ii) the understanding of natural IoBNTs (e.g., nervous nanonetwork) \cite{malak2012molecular, malak2014communication}, (iii) the development of communication and networking methods for IoBNT (e.g., molecular communications) \cite{akyildiz2019moving, lemic2021survey, akan2017fundamentals}, (iv) the design of bio/cyber and nano/macro interfaces \cite{koucheryavy2021review}, and (v) the development of new IoBNT applications \cite{balasubramaniam2013realizing, felicetti2016applications, abbasi2016nano}.

Along the aforementioned directions, this survey presents the most recent advances with respect to the theoretical foundations and practical implementation of IoBNT. To this end, we first attempt to provide a big picture of the IoBNT framework. Our discussion starts with the natural IoBNT systems, which inspire the researchers in designing artificial IoBNT systems. These include biological human-body nanonetworks, such as nervous nanonetwork, bacterial nanonetworks, and plant communication networks. Interfacing these systems with artificial IoBNT systems that monitor and control their biochemical states is expected to enable novel IoBNT applications. We extend our discussion of the IoBNT framework with the investigation of various types of BNTs, including engineered-cell based BNTs and artificial molecular and nanomachines, which ultimately determine the capabilities of IoBNT. This is followed by a review of potential IoBNT applications. Although most of them concern healthcare, there are many novel environmental and industrial applications promised by IoBNT, such as smart agriculture, food quality control, monitoring of toxic agents and pollutants, which are reviewed in this paper. 

We also provide a comprehensive review of the key technical challenges in realizing the IoBNT applications, and overview the state-of-the-art solutions and future research directions that can target them. Developing new communication methods for IoBNT is the foremost challenge, as the conventional electromagnetic (EM) techniques are either not feasible for the size- and energy-constrained BNTs or not performing well in the envisioned IoBNT application environments, such as intra-body. Molecular communications has emerged as the most promising technique to enable IoBNT, as it is already utilized by natural BNTs in a ridiculously energy-efficient and robust manner. In addition to the detailed review of MC research, we also look into other emerging communication methods proposed for IoBNT, such as those based on acoustic waves, terahertz (THz)-band EM waves, and Förster Resonance Energy Transfer (FRET). The emerging idea of using human body as an IoBNT infrastructure is also discussed through an overview of thru-body haptic communications, vagus nerve-based communication and microbiome-gut-brain-axis-based communication proposed to connect BNTs within human body.  

Bio-cyber interfaces lie at the heart of IoBNT applications, which consist in the seamless interconnection of heterogeneous technologies in diverse application environments. We provide an extensive review of electrical and optical bio-cyber interfacing technologies including biosensing-, redox-, optogenetics- and fluorescence-based techniques as well as the newly emerging magnetic and THz-based methods. IoBNT applications with high spatio-temporal resolutions in control and monitoring are expected to generate and handle significant amount of heterogeneous data, imposing critical challenges of big data processing, storage, and transfer, which are also reviewed in this paper. Self-sustaining BNTs are key to the success of IoBNT applications. Although engineered cell-based BNTs might have an inherited metabolism for energy management, artificial BNTs such as those based on nanomaterials should have dedicated mechanisms for energy harvesting (EH) and storage for continuous operation. We review various EH technologies that are suitable for the envisioned BNT architectures and IoBNT application environments. Wireless power transfer (WPT) techniques and energy storage technologies are also reviewed to provide a broader perspective on the energy challenges of IoBNT. We also discuss the security, privacy, biocompatibility and co-existence challenges of IoBNT originating from the unprecedentedly close interaction with the complex biological systems, including our own human body.

Although there are many recent survey articles focused on particular aspects of IoBNT \cite{lemic2021survey, kuscu2019transmitter, bi2021survey, akyildiz2019moving, yang2020comprehensive, akyildiz2020panacea, koucheryavy2021review, barros2021molecular}, this comprehensive review is aimed at providing a broader snapshot of the state-of-the-art in the entire IoBNT field in order to contribute to an holistic understanding of the current technological challenges and potential research directions.

\section{Framework}
\label{sec:framework}
\subsection{Natural IoBNT}
In the last several billion years, most basic single cell organisms evolved into complex systems of multi-cellular organisms composed of living nanoscale building blocks, i.e., cells, to perform the most intricate tasks in an optimized fashion. This highly coordinated structure of multicellular organisms are indeed a result of self-organized networks of cells communicating at various scales. Hence, these networks can be considered as natural IoBNTs and many lessons can be drawn in terms of effective techniques of communications and networking at nanoscale by observing the behavior of these natural IoBNTS. Here, we will describe some of the most natural IoBNTS, namely, human body nanonetworks, bacterial nanonetworks, and plant networks. 

\subsubsection{Human-body nanonetworks}
Biological systems in the human body are connected to each other and communicate primarily through molecular interactions and action potentials. These communication pathways enable the coordination of various types of cells, which are basic building blocks of life, and organization into tissues, organs, and systems with different structures and functions. The dense network of interconnected cells use signaling at various scales such as juxtracrine (signaling among cells in contact with each other), paracrine (signaling among cells in the vicinity of each other, but not in contact), or endocrine (signaling among cells  distant from each other). The performance and reliability of this intrabody networks ensures the health of the human body by preserving the equilibrium state, i.e., homeostasis, achieved by tight control of nervous system reacting to molecular and electrical inputs coming from all parts of the body and environmental cues coming through five senses. Any failure in communication in these networks will deteriorate the health and lead to diseases \cite{malak2012molecular}. For example, i) problems in electrical signaling of heart cells cause arrhythmia, i.e., irregular heartbeat, which can end in heart failure, stroke or sudden death; ii) communication problems between the brain and the body arising from the damage to protective sheath (myelin) that cover nerve fibers by immune system attacks which is the Multiple Sclerosis (MS) disease potentially causing paralysis; iii) irresponsiveness of cells to insulin which is a molecule carrying information regulating metabolism in endocrine pathways leads to diabetes; iv) irregular signaling in the microbiome-gut-brain axis, where microbes in gut and brain cells exchange information through endocrine and nervous pathways, is shown to affect mood, neurodevelopment, and obesity.   

The most advanced and complex human-body network is the nervous system \cite{malak2014communication}, composed of a very large scale network of neurons interconnected through neuro-spike \cite{balevi2013physical} and synapse \cite{khan2017diffusion} communication channels. The nervous nanonetwork distributed throughout the body transfers information about external stimuli to the brain, which is a dense network of highly complex neuron cells. The brain processes all this information and sends back commands to the body accordingly to control the vital functions, behavior and physical activities. Other networks spanning the whole body yet carrying information on a slower scale than the electrochemical pulses of nervous nanonetworks, are cardiovascular system and the endocrine system both composed of vessels carrying information in molecular form in blood and lymph, respectively \cite{malak2012molecular}.

\subsubsection{Bacterial nanonetworks}
Besides the IoBNT in multicellular organisms such as human body, single cell organisms such as bacteria show coordination and group behavior enabled by intercellular signaling. The first communication mechanism among bacteria discovered is quorum sensing (QS), the ability to detect and respond to cell population density represented by the concentration of the signaling molecules called auto-inducers \cite{waters2005quorum}. QS controls biofilm formation, virulence factor expression, production of secondary metabolites and stress adaptation mechanisms. Using this mechanism, unicellular bacteria coordinate their behavior and act as if they are a unified multicellular organism. Recently, besides the molecular means of QS, electrical means of communication among bacteria has been shown \cite{prindle2015ion}. The bacterial membrane potentials creating potassium waves through bacterial biofilms synchronize the behavior of bacteria in biofilms. In case of nutrient depletion in the center of the biofilm, this signaling mechanism warns the outer circle of bacteria in the biofilm to slow down growth allowing more nutrients to penetrate to the center.

Using the above mentioned communication mechanisms, bacteria form spatio-temporally organized community structures optimizing the growth and fitness of the whole colony which resembles a decentralized decision-making system of millions of interconnected nodes. Studies also show that bacterial colonies engage in social behavior such as competition, collaboration, and cheating during the production of public goods \cite{crespi2001evolution}. Despite the limited resources of a single bacterium, tight coordination in the bacterial populations containing sheer number of bacteria can be established. Hence, bacterial nanonetworks provide lots of clues to IoBNT researchers that are looking to form networks of large numbers of BNT devices with limited power and communication resources \cite{balasubramaniam2013multi,balasubramaniam2012opportunistic,unluturk2015genetically,unluturk2016impact}.

\subsubsection{Plant networks} 
Among the natural IoBNTs, plant networks are the most counter-intuitive since plants seem to be immobile and solitary. However, the growth and development of plants are highly dependent on the communication both within a plant itself, among different plants, and between plants and microorganisms in soil. Although there is no physical nervous system in plants, electrical communication have been observed between the roots and the body of plants \cite{awan2019information}, and capillary networks carry molecular information to the various parts of the plant along with water and nutrients. Furthermore, nearby plants use pheromone communication to coordinate their behavior to avoid overgrowth and shadowing each other and to warn each other against attacks from animals and bugs \cite{teplitski2000plants,unluturk2016end}. Considering the many species of plants in a forest, various weed and grass on top of the soil, ivies and the trees on which they live symbiotically, plant networks enable a high level coordination to share the resources and optimize growth of each plant. Another element helping this coordination is the presence of rhizobiome, i.e., the root associated microbiome, which are shaped by the plant signaling primary and secondary root metabolites \cite{olanrewaju2019plant}. In turn, the rhizobiome consisting of multiple species of bacteria helps the roots to reach necessary nutrients from the soil and protects the plant against pathogens. This rhizobiome-plant interaction significantly affects the health and growth of the plant.

\subsection{Bio-Nano Things}

In IoBNT framework, Bio-Nano Things are defined as basic structural and functional units operating at nanoscale within the biological environment \cite{akyildiz2015internet}. BNTs are expected to have typical functionalities of the embedded computing devices in IoT, such as sensing, processing, actuation, and communication.

To build BNTs, one approach is miniaturizing electrical devices with nanotechnology and encapsulating these devices for biocompatibility. However, at such a small size, miniaturized electrical BNTs suffer from lack of space for batteries to provide sufficient power and antenna generating usable frequencies. Another approach to build BNTs is utilizing biological units as substrates such as cells which can be considered standalone devices that can harvest its energy from the environment.      

% \subsubsection{Engineered cell based BNT designs} (e.g., engineered bacteria, exosomes) (BDU)

% \subsubsection{Synthetic cells as BNT} (BDU)

% %\subsubsection{Nanobiosensors and nano-stimulators }
% %(nanoimplants) (BDU, MK) \textcolor{red}{DO WE REALLY NEED THIS?}

Another important class of BNTs is molecular and nanomachines, which are tiny artificial devices with feature sizes between 1 and 100 nm, that can perform a useful task at nanoscale \cite{leigh2016genesis, lund2010molecular}. Recent years have observed the design and implementation of molecular and nanomachines with increasing complexity and sophistication, expanding the range of their applications, which now include molecular factories, self-propelling cargo carriers, nanosensors, and molecular computation \cite{ellis2018artificial}. At a coarse-grained level, molecular and nanomachines can be categorized into three main groups: molecular machines, self-assembled nanomachines, and hybrid inorganic nanomachines. 

Molecular machines are synthetic molecular systems consisting of single or a few molecules that can undergo a mechanical movement upon stimulation resulting in a useful task \cite{aprahamian2020future}. This class of BNTs can be further divided into two categories: molecular motors and switches. Molecular motors are molecular devices, typically implemented with rotaxane- or catenane-type mechanically-interlocked molecular architectures, that can perform work that in turn influences the system as a function of trajectory with chemical, light or electrochemical energy inputs \cite{kay2007synthetic, browne2010making}. Molecular motors are promising for biomimicking applications such as synthesizing new molecules from individual atoms and molecules in a programmable manner, just as their biological counterparts, such as ribosome. On the other hand, molecular switches reversibly change the state of the system upon the application of stimuli by producing no net work. Molecular switches are being utilized for high-resolution molecular sensing applications and promising for future molecular computer architectures capable of both digital and analog computing \cite{ellis2018artificial}.

Self-assembled nanomachines are nanoscale devices that are built based on autonomous or programmed organization of constituent molecules, and can perform similar functions to molecular machines, such as switching, logic gating, active propulsion, typically at larger lengths scales \cite{ellis2018artificial}. Self-assembled DNA nanomachines have particularly attracted research interest due to the high-level control of their assembly through DNA origami techniques, which enable the selective folding of DNA strands into particular designs with the use of short staple strands \cite{endo2018dna, deluca2020dynamic}. Selective targeting and versatile functionalization of DNA nanomachines have expanded their application areas, which involve bio-inspired dynamic DNA walkers, cargo-carrying DNA boxes with stimuli-responsive logic gate-based opening mechanisms, and stimuli-responsive DNA switches \cite{ramezani2020building}.

\begin{figure}
\includegraphics[width=\columnwidth]{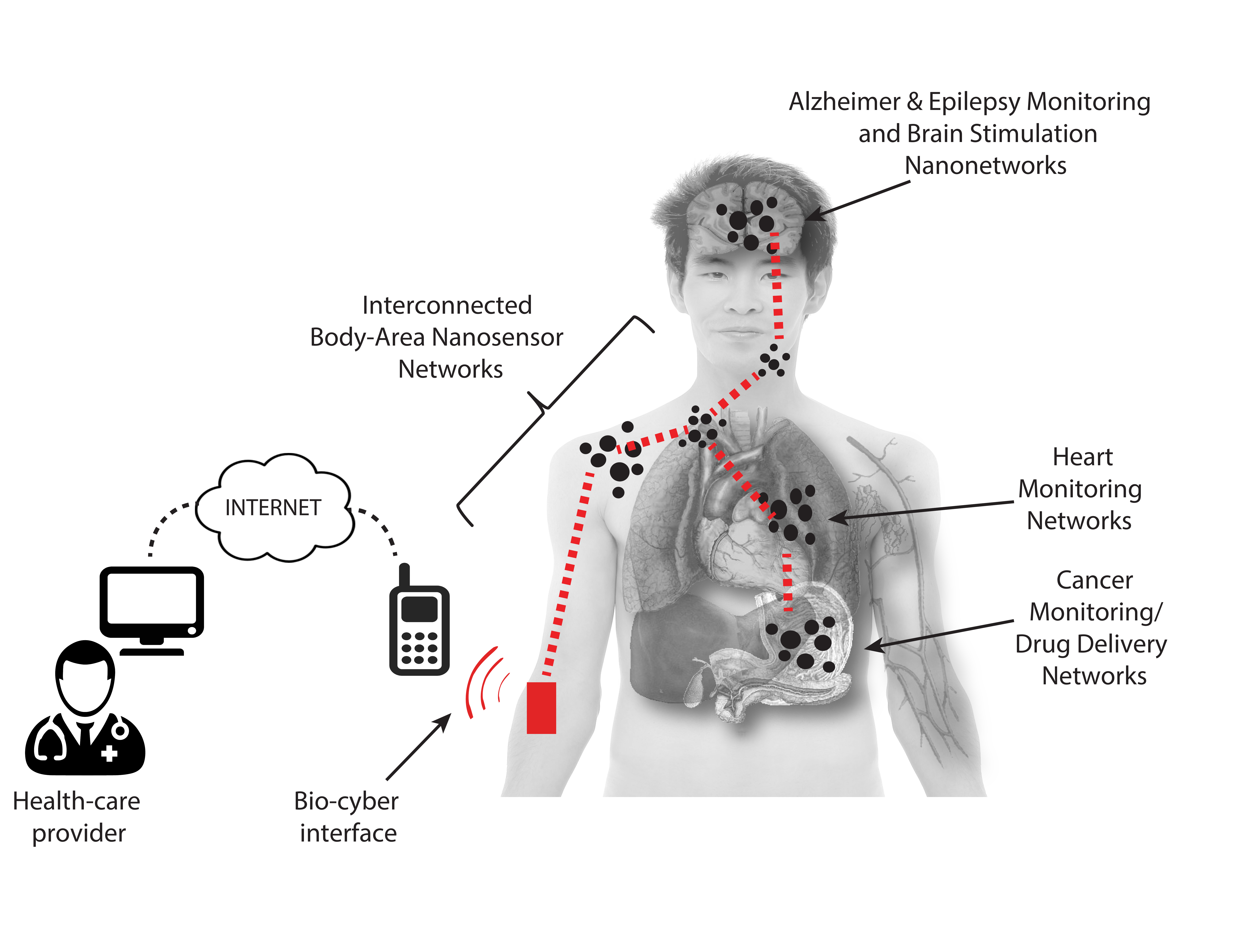}
\caption{Conceptual drawing of a continuous health monitoring application of IoBNT.}\label{fig:IoBNT} 
\end{figure}

Hybrid inorganic nanomachines are sub-100-nanometer devices that can be made of metal, metal oxide or hybrid nanoparticles (NPs) \cite{ellis2018artificial}. In comparison to self-assembled nanomachines and molecular machines, which involve soft biomolecular components, they tend to be more structurally rigid, however, less biocompatible. Additionally, their interaction with external stimuli, such as light, magnetic and electric fields, tend to be much stronger, which makes them attractive for externally controlled applications \cite{ellis2018artificial}. Janus nanomachines, which are made of Janus nanoparticles (JNPs), nanostructures with two chemically distinct parts, are the most popular inorganic nanomachines due to their anisotropic structures that give rise to exceptional propulsion capabilities \cite{sanchez2015chemically}. This anisotropy results in a chemical potential or thermal gradient in JNPs upon chemical catalysis or external light irradiation, which in turn, leads to the phoretic flow of surrounding fluid around the entire JNP surface. As a result of this phoretic flow, JNPs actively move in the opposite direction. There are also proposed architectures of self-propelled Janus nanomotors working based on the decomposition of hydrogen peroxide into oxygen as a driving force. Moreover, the use of mesoporous silica nanoparticles (MSNs) as JNPs has opened up new biomedical opportunities which involve the targeted delivery and controlled release of therapeutic and diagnostic agents encapsulated in their porous structure \cite{xuan2014self}. 

Although much has been done to devise exquisite and complex molecular and nanomachine architectures that can perform sensing, cargo transport, and switching operations, the potential of interconnecting these tiny machines for a wider range of biomedical applications has only recently attracted significant attention. Several communication modalities have already been considered to enable controlled interaction and coordination of these devices, such as diffusion-mediated communication, which is based on the exchange of small molecules, e.g., glucose, through the signaling cascades triggering enzymatic reactions that fuel the movement of the receiver devices, and cell or cell-free genetic circuits that trigger the expression of a certain kind of protein, e.g., green fluorescent protein (GFP), in the receiver BNTs \cite{luan2020leveraging, de2021chemical, chen2018bioinspired, llopis2017interactive}. External energy-mediated communication has also been widely studied to enable the small networks of molecular and nanomachines. This form of interaction occurs through several biophysical phenomena, such as pore formation and modulation of enzyme cascade reactions, which are triggered by external stimuli such as light, chemicals, temperature, and electric and magnetic fields, and provides higher level of spatiotemporal control compared to the diffusion-mediated communication \cite{luan2020leveraging}. Non-covalent interactions considered for molecular and nanomachines involve the short-range electrostatic and hydrophobic/hydrophilic interactions, as well as complex formation through reversible ligand-receptor binding interactions. Lastly, inducing dynamic collective behaviors of active nanomachines, such as Janus nanomachines, through external stimuli, e.g., light, electric and magnetic fields, has also attracted great attention due to their emergent out-of-equilibrium properties resembling natural systems \cite{wang2020coordinated, yan2016reconfiguring, dey2019dynamic}. Incorporation of these interacting molecular and nanomachines into the larger IoBNT framework as heterogeneous BNTs can enable unprecedented therapeutic and diagnostic applications via exquisite external, distributed, or programmed control with high spatiotemporal resolution. 

\subsection{IoBNT Applications}
The IoBNT will enable a plethora of applications in many fields where the connection of biological entities and nanodevices to the Internet leads to unprecedented ways of interfacing with biology due to IoBNT's inherent biocompatibility, reduced invasiveness, and low power consumption. In the rest of this section, we discuss the potential of IoBNT in biomedical applications, smart agriculture, and environmental applications.  

\subsubsection{Biomedical Applications}

The most promising applications of IoBNT are envisioned to be in the biomedical field where IoBNT would play a crucial role in healthcare. IoBNT comprising nanonetworks of biosensors and actuators operating near, on, or in the body, will enable real-time remote monitoring and control of patients' health. 

A nanosensor network deployed in cardiovascular system monitoring vital signs such as heart rate, blood pressure, EEG signals, and blood oxygen and carbon dioxide levels may reveal abruptly occurring diseases such as heart attack and automatically alert healthcare providers. Meanwhile, continuous long-term monitoring of these vital signs may be used for management of chronic diseases as well as data collection to predict future attacks. 

Recently, researchers also considered applying IoBNT concept for detection and mitigation of infectious diseases \cite{akyildiz2020panacea} where bio-hybrid BNTs constantly monitors for biomarkers released by infectious microorganisms. Other biomarkers that would be interesting to monitor by IoBNT would be glucose. A sudden changes in glucose levels can be deadly for diabetic patients, IoBNT can alert the patient against low/high glucose levels and can help adjust precise and timely administration of insulin automatically. Similarly, IoBNTs can be used for hormonal therapy management in cancer treatments or hormone replacement therapies in sex change \cite{akan2016fundamentals}.     

Besides monitoring applications, IoBNT can also lead the realization of next generation smart drug delivery applications. To spare the non-target organs and tissues from the side effects of drugs, BNTs can deliver medicine to targeted regions in human body. BNTs encapsulating drug molecules can either actively search for or be directed externally to target cells and release the drugs only on target location.

\subsubsection{Smart Agriculture}
Humans are not the only organisms that can benefit from remote health monitoring with IoBNT. The health of animals such as cattle and poultry can be also interrogated by IoBNT to ensure the health of the animals and the quality of their products such as meat, milk, and eggs. Another benefit of IoBNT to agriculture would be through monitoring of plants by measuring their health through BNTs deployed on the plants or in the soil. This can be also supported by BNTs monitoring and controlling smart irrigation systems, actively fertilizing the soil, and deterring bugs and wildlife damaging crops.

%  \subsubsection{ Biocomputing}
%   ultra-dense data storage with DNA, high-rate data transfer with bacteria

%  \subsubsection{Covert communication systems}

%  \subsubsection{Food safety and quality monitoring}

\subsubsection{Environmental Applications}
Another promising area for IoBNT applications is environmental monitoring. By deploying IoBNT networks in water supply and distribution systems, it might be possible to detect pollutants in the water and use nano-filters to remove harmful substances and toxic agents contained in it. A similar system can be deployed to combat air pollution in crowded cities. Another environmental application can be listed as handling the growing problem of waste management where IoBNTs can be used to sort and process waste. Nanosensors can sense and tag different materials and nanoactuators can biodegrade the tagged materials or alert service providers to remove potentially toxic waste that might pollute water or soil.  

\section{Challenges}

\subsection{Communication Methods for IoBNT}

Conventional forms of electromagnetic (EM) communications are deemed not suitable for connecting BNTs, mainly due to the antenna size limitations, biocompatibility concerns, and the severe attenuation of EM signals in physiological media relevant for IoBNT applications \cite{akyildiz2015internet}. Because of these challenges, researchers have started a quest for alternative communication methods to extend our connectivity to nanoscales. We can classify the proposed nanocommunication methods into two main types: (i) Molecular communications (MC), (ii) THz-band EM. Other techniques based on magnetic coupling, F{\"o}rster Resonance Energy Transfer (FRET), heat transfer and acoustic energy transfer have also been proposed for nanonetworks. In the rest of this section, these techniques will be briefly overviewed, with a particular focus on MC, which is considered as the most promising nanocommunication method to enable IoBNT.

\subsubsection{\textbf{Molecular Communications}} 
Molecular communications is a bio-inspired communication technique, that uses molecules to transfer information. More specifically, a physically distinguishable feature of molecules, such as their type and concentration, is used to encode information, and random molecular motion in a fluidic channel is exploited as a means of signal propagation for information transfer. MC is radically different from conventional communication paradigms, e.g., EM communications, in various aspects such as the size and type of network entities, information transmission mechanisms, noise sources and fundamental performance limits including transmission delay, achievable data rates, coverage and power consumption.

\begin{figure}
\includegraphics[width=\columnwidth]{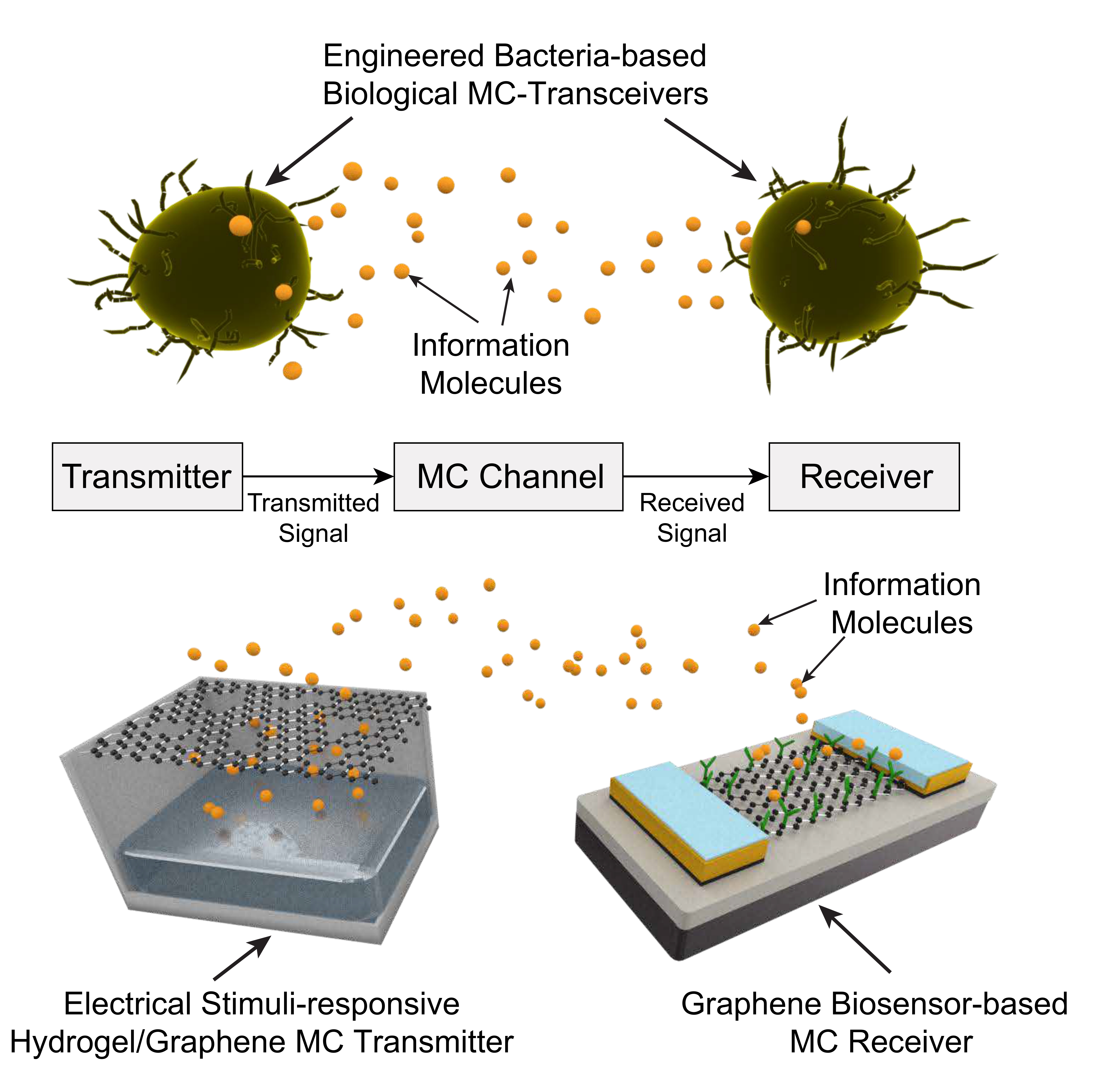}
\caption{Components of an MC system with biological and nanomaterial-based MC transmitter and receiver design approaches. }\label{fig:MolCom} 
\end{figure}

Example MC scenarios between pairs of nanomachines are depicted in Fig. \ref{fig:MolCom}, where the messages are encoded into the concentration of molecules, and then transmitted to the receiver via molecular propagation in a fluidic channel. The information can also be encoded into the type, release time, or the electronic state of the molecules \cite{akan2017fundamentals}. Different kinds of propagation methods for molecular messages are investigated in the literature, such as passive diffusion, active transport with molecular motors  \cite{moore2006design}, convection, and transport through gap junctions \cite{kilinc2013information}. Among these, passive diffusion is the most promising, as it does not require energy consumption, and thus perfectly suits the energy limitations of the envisioned nanomachines. 

% \begin{figure*}[!t]
% 	\centering
% 	\includegraphics[width=10cm]{Figs/IoBNT_MC}
% 	\caption{Components of an MC system with biological and nanomaterial-based MC transmitter and receiver design approaches. }
% 	\label{fig:MolCom}
% \end{figure*}

MC channel has many peculiar characteristics. For example, the discrete nature of information carriers, i.e., molecules, results in molecular counting noise, which is of similar nature with the shot noise occurring in photonic devices \cite{pierobon2011diffusion}. The stochastic nature of the ligand-receptor binding process occurring at the receiver gives rise to colored noise, also leading to a strong correlation between molecular propagation process and reception \cite{pierobon2011noise}. The slow nature of diffusion leads to a substantial amount of channel memory, which in turn, causes severe intersymbol interference (ISI), and limits the achievable data transmission rates \cite{pierobon2012intersymbol}. The same reason also causes a significant delay in the transmission \cite{unluturk2013rate}.

Deviations from the conventional means of communications necessitate radically different ideas for the design of transmitter and receiver architectures, and communication techniques for MC, and new approaches to channel modeling.  

{\textbf{a) Transmitter and Receiver Architectures for MC:}}
\label{sec:TransmitterReceiverMC}
% should be revised by referring to the BNT section... 
There are mainly two design approaches considered for artificial nanomachines that can perform MC and form MC nanonetworks within the IoBNT framework. The first approach is to build the components of nanomachines using newly discovered nanomaterials, such as two-dimensional graphene, and one-dimensional silicon nanowire (SiNW) and carbon nanotube (CNT),  which all manifest extraordinary characteristics at the interface of biology and electronics \cite{kuscu2016physical}. The other approach relies on synthetic biology, and envisions the use of engineered, i.e., genetically modified, bacteria as artificial nanomachines with communication functionalities wired into their intracellular signaling networks \cite{unluturk2015genetically}.

The physical nature of the BNTs determines the potential transmitter and receiver architectures. The MC transmitter of a BNT should perform the modulation of MC signals, and the release of molecules into the channel upon a stimulation by an external source, or as a result of an internal biochemical or electrical process. The receiver of a BNT is responsible for detecting the incoming molecular messages, transducing them into a processable signal, and extracting the encoded information through signal processing. The decoded information can then be used by the BNT to perform a prescribed operation, e.g., modulation of gene expression or translocation. Therefore, the performance of the transmitter and receiver is critical for the proper operation of a BNT within an IoBNT application.

Nanomaterial-based design approaches for MC transmitter mainly draw on the existing drug delivery technologies, such as stimuli-responsive hydrogels, molecule release rate of which is controlled by an electrical or chemical stimuli. Synthetic biology-based approaches, on the other hand, rely on making use of the already existing molecule release mechanisms of living cells, and tailoring these functionalities through genetic modifications to realize the desired MC modulation schemes. There are also theoretical MC transmitter designs that exploit stimuli-responsive ion channels to trigger the release of molecules in an externally controllable fashion \cite{soldner2020survey}.

Nanomaterial-based receiver designs are widely inspired by nanobiosensors, which share a common objective with MC receivers, that is to transduce biomolecular signals into a signal form suitable for processing. Although there are many nanobiosensor designs differing in their transducing mechanisms and the resulting signal form at the output, field-effect-transistor (FET)-based nanobiosensors have attracted the most attention for MC receiver design due to their scalability, simple design similar to conventional FETs, internal signal amplification by electrical field-effect, label-free operation, and the electrical output signals that allow fast processing of received signals. More importantly, FET-based nanobiosensors provide a wide range of design options. For example, they can accommodate different types of nanomaterials, e.g., graphene, SiNW, CNT, as the transducer channel, the conductivity of which is modulated by the molecular concentration in its proximity through the alteration of the surface potential and electrical field-effect. FET-based nanobiosensors have also a biorecognition layer, which replaces the gate electrode in conventional FETs, and consists of receptor molecules selectively binding target molecules via affinity-based ligand-receptor interactions. Depending on the transducer channel material, the biorecongition layer can host a wide range of receptor molecules, ranging from proteins to DNAs. 

Among other options, graphene FET (GFET) biosensors provide unique advantages for the practical design of MC receiver. The main advantage of graphene is its high sensitivity to the charged analytes, e.g., proteins and DNAs, due to its extremely high carrier mobility and one-atom thickness, exposing all its atoms to the environment. The advent of new types of receptors, e.g., aptamers, has broadened the target range of nanoscale FET biosensors from ions to proteins, peptides, and even whole cells. Aptamers are short functional oligonucleotides (typically 20-60 nucleotides). Their base sequences for specific targets are identified from an oligonucleotide library with an in vitro process called systematic evolution of ligands by exponential enrichment (SELEX). Their application in biosensors has gained momentum due to their wide target range, chemical stability, and ease of production. Combined with the exceptional properties of graphene and aptamers, the ability of nanoscale FET biosensors to provide selective, label-free and continuous detection makes GFET aptamer-based biosensors, i.e., GFET aptasensors, very promising candidates for the design of MC receiver.

Biological MC receiver designs are based on the enhancement of biosensing and biochemical signal processing functionalities of livings cells with synthetic biology tools for the receiver operation. This approach consists in the design of new synthetic receptors that can provide more sensitivity and selectivity in physiological environments, for example, through kinetic proof reading mechanisms \cite{kuscu2019channel}, and the implementation of new chemical reaction networks within the cell that can realize the required computations for decoding the received MC signals. Synthetic biology is already mature enough to allow performing complex digital computations, e.g., with networks of genetic NAND and NOR gates, as well as analog computations, such as logarithmically linear addition, ratiometric and power-law computations, in synthetic cells \cite{daniel2013synthetic}. Synthetic gene networks integrating computation and memory is also proven feasible \cite{purcell2014synthetic}. More importantly, the technology enables implementing BNTs capable of observing individual receptors, as naturally done by living cells. Hence, it stands as a suitable domain for practically implementing more information-efficient MC detectors based on the binding state history of individual receptors. 

{\textbf{b) MC Channel Modeling:}}
To design effective and efficient MC systems addressing the needs of the envisioned IoBNT applications, it is important to have a theoretical framework which can be used to optimize the physical components of the system with ICT performance metrics. Because of this, there has been tremendous interest in modeling the MC channels to find the ultimate performance limits in terms of information theoretical capacity and data rate. In majority of studies, MC channel is usually assumed to be unbounded where information-carrying molecules propagate through free diffusion with the underlying phenomenon of Brownian motion \cite{llatser2011diffusion, pierobon2013capacity}. In a few studies, diffusion is accompanied by a flow which directs the propagating molecules to a distant receiver \cite{kuscu2018modeling, noel2014optimal, bicen2013system}, whereas some studies also consider the existence of reactive molecules within the channel which can chemically degrade the information carrier molecules and reduce the intersymbol interference. A few studies consider bounded MC channels, for example microfluidic channels where molecules propagate through convection-diffusion. In majority of these studies, it is assumed that the molecules are transmitted from a hypothetical point source, which is capable of releasing a known number of molecules to the channel in the form of an impulse signal at a given time instant. On the other hand, the receiver is typically assumed to be a transparent instrument, which is capable of counting every single molecule in a hypothetically defined space \cite{noel2014optimal}, or an ideal absorbing instrument capable of counting each molecule that is absorbed \cite{yilmaz20143}. Common to these studies is the ignorance of the impact of the physical architectures of the transmitter and receiver on the communication channel. As such, researchers have been able to adopt the EM-inspired simplifications in modeling, such as linear and time-invariant (LTI) channel characteristics with additive white Gaussian noise, neglecting the effects of interactions and correlations resulting from transmitter and receiver architectures and channel geometry. This leads to a serious discrepancy between theory and practice, as revealed by the initial MC experiments performed with ‘macroscale’ sensors and dispensers utilized as MC transmitter and receiver, respectively, showing that the nonlinearity and time-variance caused by the operation of transmitter and receiver invalidate the models built upon these assumptions \cite{farsad2013tabletop, unterweger2018experimental}.

On the other hand, some research groups have studied MC receivers that rely on ligand-receptor binding reactions, the common molecular sensing method in natural MC \cite{deng2015modeling, kuscu2016modeling}. Deterministic models, assuming free diffusion and point transmitter, have been developed for a virtual MC receiver with ligand receptors. Although the consideration of ligand receptors has advanced the accuracy of the models one step further, the employed assumptions about the transmitter and channel strictly limit the applicability of these models. Additionally, stochastic receiver models are developed for FET nanobiosensor-based MC receivers \cite{kuscu2016modeling}. In \cite{kuscu2018modeling}, a model for MC with 2D biosensor-based receivers in microfluidic channels is provided. However, these initial models also rely on unrealistic assumptions, e.g., equilibrium conditions in ligand-receptor binding reaction, and ignore the implications of the receiver geometry.

Therefore, there is still a need for a bottom-up physical modeling approach originating from first principles, capturing all interactions in practical transmitter, channel and receiver architectures, causing nonlinear and time-varying behavior and unconventional noise and interference that may have decisive impacts on the development of MC techniques for IoBNT. 

{\textbf{c) Experimental MC System Testbeds and Practical Demonstrations:}}
Testbeds are crucial for validating theoretical models, and practically evaluating the performance of new communication techniques. However, research for building experimental MC systems has just recently gained momentum. Few studies in MC literature have focused on `macroscale' implementation of MC systems with off-the-shelf components. For example, in \cite{farsad2013tabletop, koo2016molecular}, the isopropyl alcohol (IPA) is used as airborne information carriers, and commercially available metal oxide semiconductor alcohol sensors are used as the MC receiver. The transmission of molecules is realized by electrically-controlled spray nozzles. In \cite{farsad2017novel}, the information is encoded in pH level of the transmitted fluid instead of molecular concentration or type. The acidic and basic fluids are injected into off-the-shelf flexible tubes via peristaltic pumps, and a macroscale pH meter is used as the MC receiver. In another testbed \cite{unterweger2018experimental}, magnetic nanoparticles (MNs) are employed as information carriers, which are injected into off-the-shelf mm-scale flexible tubes by flow pumps, and propagate through convection and diffusion. In this study, for the detection of messages encoded into MN concentration, the authors designed bulky detector coils placed around the tubes coupled with capacitors to form a resonator circuit, which informs about the concentration through a change in inductance and shift in resonance frequency. In this implementation, the designed receiver acts only as an observer, and does not physically interact with the information carrier molecules. The focus of the aforementioned studies is on macroscale MC using commercially available channels, and off-the-shelf sensors or bulky detectors as receivers that are not physically relevant for the application domain of MC and IoBNT.

Recently, the first micro/nanoscale demonstration of an MC system is reported in \cite{kuscu2020graphene}. In this study, the authors provide the results of MC experiments using a custom-made microfluidic testbed with a graphene FET DNA biosensor-based MC receiver integrated into a microfluidic channel. A commercially available microfluidic flow control system is used to pump single-stranded DNA (ssDNA) molecules of different molecular concentrations into the microfluidic channel. Graphene transducer channel of the receiver functionalized with complementary ssDNAs transduces the real-time concentration of the propagating DNA molecules into electrical signals, which are then used for detection. The authors of the study report nM-level sensitivity and single-base-pair-mismatch selectivity for the receiver. However, they also note the very low communication rates on the order of 1 bit/minute, mainly resulting from the slow association-dissociation kinetics of DNA hybridization.

Biological MC testbeds have also been reported by many research groups. For example, in \cite{krishnaswamy2013time} and \cite{bicen2015efficient}, authors implement a microfluidic MC testbed with genetically engineered Escherichia coli (E. coli) bacteria acting as receiver nanomachines. The bacteria in these studies have been engineered to respond to certain biomolecules, e.g., N-(3-Oxyhexanoyl)-L-homoserine lactone (C6-HSL) and N-Acyl homoserine lactone (AHL), by expressing green fluorescent proteins (GFP), which can later be detected via fluorescent microscopy upon excitation with light of certain wavelengths. Both studies report extremely low communication rates on the order of 1 bit/hour due to the lengthy process of gene expression required for each bit transmission. In \cite{grebenstein2019molecular}, the authors prefer a different approach by exploiting the light-responsive proton pump gloeorhodopsin (GR) located in the bacterial membrane to obtain an optically controlled MC transmitter that can export protons into the fluidic channel upon the application of external light stimuli. Accordingly, protons are used as information carriers, which propagate through structural diffusion in water, and are detected by a pH sensor acting as the receiver. Using this testbed, the authors report communication rates on the order of 1 bit/minute. Although biological designs have been demonstrated individually for both MC transmitter and receiver, there is yet to be any practical implementation of an entirely biological testbed for end-to-end MC. 

{\textbf{d) Development of MC Techniques:}}
The unconventional characteristics of MC, such as discrete nature of information carriers and slow nature of propagation mechanisms, which bear no similarity to conventional EM communications, lead to various challenges, such as high channel memory causing severe ISI, non-Gaussian noise sources, time-variance, and very low communication rates, as revealed by several theoretical investigations \cite{kuscu2018modeling, kuscu2019transmitter}. The initial experimental studies performed on both macro- and micro-scales also demonstrated the high level of nonlinearity mainly resulting from the characteristics of sensors utilized as receivers \cite{farsad2013tabletop, koo2016molecular, kuscu2020graphene}. One can expect that practical MC system implementations for IoBNT applications may face many more challenges, such as molecular interference due to existence of different types of molecules in the channel, environmental fluctuations, such as those in flow velocity and temperature, ionic screening in physiologically relevant environments preventing the receiver from detecting the electrical charges of information molecules, and new noise sources such as electronic 1/f noise in nanomaterial-based MC receivers. Therefore, MC requires new communication methods that account for these peculiarities, and overcome their detrimental effects on the communication performance. Considering physical limitations of the envisioned BNTs, these techniques should be also low-complexity and low-energy, i.e., low-molecule-use. 

We summarize some of the major problems stemming from the limitations associated with the physical properties of the MC channel, transmitter and receiver architectures as follows: 
\begin{itemize}
\item \textbf{Intersymbol Interference (ISI):} Due to the slow nature of molecular diffusion in MC channel, severe ISI occurs in both forward and backward direction, which is the main factor limiting the communication rate. The effect of ISI is less pronounced in flow-based MC channels; however, the slow reaction kinetics at the receiver surface might compound the ISI, as revealed in \cite{kuscu2020graphene}. Therefore, MC techniques should account for ISI, either removing it or reducing its effects.
\item \textbf{Nonlinearity and Time-variance:} The nonlinearity of the MC system results from the nonlinear transmission and propagation dynamics, and the reaction-based receiver mechanisms. On the receiver side, in particular, the saturation of the receiver could have substantial effect on the detection performance. Therefore, the developed modulation and detection techniques should account for nonlinearity. Time-variance can result from the fluctuations in the flow conditions, as well as from the time-varying molecular interference level in the channel. 
\item \textbf{Molecular Interference:} The existence of other molecules in the MC channel can originate from an irrelevant biological process, or another MC system coexisting in the same channel. The interference manifests itself on the receiver side, as the selectivity of receptors against information molecules is far from ideal in practice, and thus, many different types of molecules having finite affinity with the receptors, could also bind the same receptors, resulting in considerable interference at the received signal. To overcome this problem, new detection methods exploiting the frequency-domain characteristics of the receiver reaction and transducing processes can be developed to increase the selectivity \cite{kuscu2019channel}. Moreover, the receptor cross-talk resulting from multiple types of molecules can be exploited to develop new modulation techniques to boost the communication rate. 
\item	\textbf{Noise:} In addition to particle counting noise and ligand-receptor binding noise, which are well investigated in the MC literature, the physical architecture of the receiver can lead to new noise sources. For example, in nanomaterial-based designs, thermal noise and electronic noise, e.g., 1/f noise, of the receiver can be expected to severely undermine the reliability of communication.  
\item	\textbf{Ionic Screening:} One of the main problems particularly observed at FET biosensor-based receivers is the ionic screening in physiologically relevant fluids, which decrease the SNR tremendously. The ions in the channel fluid can cause the screening of electrical charges of information molecules, resulting in reduced  effective charge per molecule that can be detected by the receiver via field-effect. The strength of ionic screening depends exponentially on the distance of bound information molecules from the surface of the receiver's transducer channel. Numerous solutions exist in the biosensing literature that partially overcome this widely-observed problem. For example, using small-size receptors, e.g., aptamers, can allow the bound information molecules to approach the receiver surface, increasing their effective charge \cite{ohno2010label, elnathan2012biorecognition}. Alternatively, high-frequency AC biasing at the receiver, exploiting the oscillating dipole moments of the bound information molecules, can be employed to overcome the ionic strength in exchange for increased complexity on the receiver side \cite{kulkarni2012detection}.   
\item	\textbf{Low Communication Rate:} Slow diffusion and reaction kinetics of molecules might result in very low-communication rates, as shown in some of the recent practical MC demonstrations. These physical limitations call for new modulation and detection techniques that simultaneously exploit multiple properties of molecules, e.g., concentration and type, to boost the communication rate for MC systems. 
\end{itemize}

Modulation techniques in MC fundamentally differ from that in conventional EM communications, as the modulated entities, i.e., molecules, are discrete in nature, and the developed techniques should be robust against highly time-varying characteristics of the MC channel, as well as inherently slow nature of the propagation mechanisms \cite{akan2017fundamentals}. Exploiting the observable characteristics of molecules, researchers have proposed to encode information into the concentration, type, or release time of the molecules \cite{kuscu2019transmitter, kuran2011modulation}. The simplest modulation method proposed for MC is on-off keying (OOK) modulation, where a binary symbol is represented by releasing a number of molecules or not releasing any \cite{mahfuz2010characterization}. Similarly, using a single type of molecule, concentration shift keying (CSK), that is analogous to amplitude shift keying (ASK) in traditional wireless channels, is introduced in order to increase the number of transmitted symbols by encoding information into molecular concentration levels \cite{kuran2012interference}. Molecular information can also be encoded into the type of molecules, i.e., molecule shift keying (MoSK) \cite{kuran2011modulation}, or into both the type and the concentration of molecules to boost the data rate \cite{arjmandi2013diffusion}. Additionally, the release order of different types of molecules \cite{atakan2012nanoscale}, and the release time of single type of molecules \cite{garralda2011diffusion} can be modulated to encode information in MC. Finally, in \cite{kim2013novel}, authors propose the isomer-based ratio shift keying (IRSK), where the information is encoded into the ratio of different types of isomers in a molecule, i.e., molecule ratio-keying.

To overcome the noisy and ISI-susceptible nature of MC channels, several channel coding techniques which are adopted from EM communications, e.g., block and convolution codes, or developed specifically for MC, such as the ISI-free coding scheme employing distinguishable molecule types, have been studied. Detection is by far the most studied aspect of MC in the literature. Several methods varying in complexity have been proposed to cope with the ISI, noise, and even the nonlinearity of the channel, such as optimal Maximum Likelihood (ML)/Maximum A Posteriori (MAP) detection methods, noncoherent detection, sequence detectors with Viterbi algorithm \cite{llatser2013detection, kilinc2013receiver, noel2014optimal, li2015low, kuscu2019transmitter}. Synchronization problem is addressed by both developing self-synchronizing modulation techniques and asynchronous detection methods. However, these methods are developed based on existing theoretical models of MC, which largely lack physical correspondence. Therefore, the performance of the proposed methods is not validated, which poses a major problem before practical MC systems and IoBNT applications.

\subsubsection{\textbf{Human Body as IoBNT Infrastructure}}
MC can typically support only very low communication rates due to the slow diffusion dynamics of molecules. Moreover, MC is prone to errors because of high level of noise and molecular interference in crowded physiological media, as well as due to attenuation of molecular signals as a result of degradation via various biochemical processes, making it reliable only at very short ranges \cite{akan2017fundamentals}. 

On the other hand, human body has a large-scale complex communication network of neurons extending to various parts of the body and connecting different body parts with each other through electrical and chemical signaling modalities \cite{malak2012molecular}. A part of the nervous system also senses external stimuli via sensory receptors and transmits the sensed information to the central nervous system, where a reaction is decided \cite{malak2014communication}. In that regard, the nervous system provides a ready infrastructure that can potentially connect nanomachines in distant parts of the body with each other and with the external devices. In fact, there are many proposal in this direction that both theoretically and experimentally investigate the idea of using the nervous system as an IoBNT backbone inside human body. 

In \cite{hanisch2017digital}, authors consider a thru-body haptic communication system, where the information encoded into tactile stimulation is transmitted to the brain through the nervous system, resulting in a discernible brain activity which is detected by ElectroEncephaloGraphy (EEG) and used to decode the transmitted information. An analytical framework based on the computational neuroscience models of generation and propagation of somatosensory stimulation from skin mechanoreceptors is developed for the analysis of the achievable data rate on this communication system. Authors show that the system can support bit rates of $30$-$40$ bit per second (bps) employing an OOK modulation of tactile stimulation taps at the index finger.

In \cite{abbasi2018controlled}, the authors practically demonstrate a controlled information transfer through the nervous system of a common earthworm, which stands as a simple model system for bilaterian animals including humans. In the demonstrated setup, authors use external macroscale electrodes to interface with the earthworm's nervous system. Accordingly, the stimulation carrying the encoded information is applied at one end of the nerve cord, and the resulting nerve spikes are recorded at the other end. Through the application of different modulation schemes, e.g., OOK, frequency shift keying (FSK), the authors demonstrate data rates up to $66.61$ bps with $6.8\times10^{-3}$ bit error rate. 

In \cite{donohoe2020deep}, the authors propose to use vagus nerve to deliver instructions to an implanted drug delivery device near the brainstem via compound action potentials (CAP) generated by the application of electrical impulses at the neck, known in the literature as the vagus nerve stimulation (VNS). Applying an OOK modulation, the authors theoretically show that the vagus nerve can support data rates up to $200$ bps and unidirectional transmission ranges between $60$ mm and $100$ mm, which is promising for enabling the communication of distant BNTs at a rate that is much higher than the typical MC rates. 

A different approach to make use of the natural human body networks for IoBNT is investigated in \cite{akyildiz2019microbiome}, where authors propose to use Microbiome-Gut-Brain-Axis (MGBA) to connect distant BNTs. MGBA is a large scale heterogeneous intrabody communication system composed of the gut microbial community, the gut tissues, and the enteric nervous system. In MGBA,  a bidirectional communication between the central nervous system and the enteric nervous system surrounding the gastrointestinal track (GI track) is realized via the transduction of electrical signals in the nervous system into molecular signals in the GI track, and vice versa. The axis has recently attracted significant research interest due to the discoveries underlining the relation of MGBA signaling with some neurological and gut disorders such as depression and irritable bowel syndrome (IBS). In the research roadmap proposed in \cite{akyildiz2019microbiome}, the authors envision BNTs as electrical biomedical devices, e.g., cardiac pacemaker, brain implants, insulin pumps, and biological devices, e.g., synthetic gut microbes and artificial organs, interconnected through the MGBA. They also investigate the possibility of a link between the IoBNT and the external environment via molecular (alimentary canal) and electrical (wireless data transfer through skin) interfaces. 

\subsubsection{\textbf{Other Nanocommunication Modalities for IoBNT}}

{\textbf{a) THz-band Electromagnetic Nanocommunication:}}
Conventional electromagnetic (EM) communication is not deemed suitable for IoBNT because the size of BNTs would demand extremely high operating frequencies \cite{lemic2019survey}. Fortunately, graphene-based nanoantennas based on surface plasmon polariton (SPP) waves have been shown to support frequencies down to 0.1 THz, much lower than their metallic counterparts, promising for the development of high-bandwidth EM nanonetworks of nanomaterial-based BNTs using the unutilized THz-band (0.1-10 THz) \cite{jornet2013fundamentals}. In this direction, several plasmonic transceiver antenna designs using graphene and related nanomaterials (e.g., CNT), whose properties can be tuned by material doping and electric field, have been investigated \cite{llatser2012graphene, da2009carbon}. However, several challenges exist for the practical implementation of THz-band nanonetworks, such as the very limited communication range resulting from high propagation losses due to molecular absorption, and low transmission power of resource-limited nanodevices. These challenges are being addressed by developing new very-short-pulse-based modulation schemes to overcome the limitations of THz transceivers in terms of power \cite{jornet2014femtosecond, mabed2017enhanced}, and designing directional antennas and dynamic beamforming antenna arrays to overcome the propagation losses \cite{singh2020design}. High density of BNTs in envisioned IoBNT applications also pose challenges regarding the use of the limited spectrum, which are addressed by new medium access protocols for dense THz nanonetworks \cite{mabed2018flexible, shrestha2016enhanced}. 

{\textbf{b) Acoustic Nanocommunication:}}
Ultrasonic nanocommunication has also been considered for connecting robotic BNTs inside the fluidic environment of human body due to its well-known advantages over its RF counterpart in underwater applications \cite{hogg2012acoustic, santagati2013ultrasonic, santagati2014medium}. In \cite{hogg2012acoustic}, it is shown that the best trade-off between efficient acoustic generation and attenuation is realized when the acoustic frequency is between $10$ MHz and $300$ MHz for distances around $100$ $\mu$m. The authors also show that the power harvested from ambient oxygen and glucose can be sufficient to support communication rates up to $10^4$ bps. In \cite{santagati2016experimental}, the authors provide a testbed design for ultrasonic intrabody communications with tissue-mimicking materials and as a result of extensive experiments, they report communication rates up to $700$ kbps with a BER less than $10^{-6}$. 

An alternative approach proposed in \cite{santagati2014opto} suggests the use of optoacoustic effect for the generation and detection of ultrasonic waves via a laser and an optical resonator, respectively. It is shown that optoacoustic transduction brings multiple advantages for ultrasonic nanocommunications, such as higher miniaturization, bandwidth and sensitivity over traditional piezoelectric/capacitive transduction methods. 

{\textbf{c) FRET-based Nanocommunication:}}
Single molecular BNTs are not capable of performing active communications, as in the case of MC and THz-band EM communications. On the other hand, external stimuli can supply the necessary means of information transfer. One such method is based on F{\"o}rster Resonance Energy Transfer (FRET), which is a non-radiative and high-rate energy transfer between fluorescent molecules, such as fluorescent proteins and quantum dots (QDs) \cite{kuscu2011physical}. The method requires an external optical source for the initial excitation of donor molecules, which then transfer their energy to ground-state acceptor molecules in their close proximity. Encoding information into the excited state of molecules, short-range (5-10 nm) but very high-rate (on the order of Mbps) information transfer can be realized by this method \cite{kuscu2015internet}. Additionally, bioluminescent molecules can be utilized as donors that are excited upon binding specific target molecules, promising for single molecular sensor networks within an IoBNT application \cite{kuscu2014communication, kuscu2014coverage}. It is shown that the limited range of FRET-based nanocommunication can be extended to 10s of nanometers by multi-step energy transfer processes and multi-excitation of donor molecules  \cite{kuscu2013multi, kuscu2014fret}. Lastly, an experimental study demonstrated a high rate data transfer (250 kbps with a BER below $2\times10^{-5}$) between fluorescent-dye nanoantennas in a MIMO configuration \cite{kuscu2015fluorescent}. 

\subsection{Bio-Cyber and Nano-Macro Interfaces}

Most of the envisioned IoBNT applications require a bidirectional nano-macro interface that can seamlessly connect the intrabody nanonetworks to the external macroscale networks, and vice versa \cite{dressler2015connecting, civas2021universal}. Considering that the MC is the most promising method for intrabody IoBNT, the interface should be capable of performing the conversion between biochemical signals and other signal forms that can be easily processed and communicated over conventional networks, such as electromagnetic, electrical, and optical. Several techniques are considered for enabling such a nano-macro interface. 

\subsubsection{\textbf{Electrical Interfaces}} 

These are the devices that can transduce molecular signals into electrical signals, and vice versa. Electrical biosensors can readily serve the function of converting MC signals into electrical signals (see Section \ref{sec:TransmitterReceiverMC} for the use of electrical biosensors as MC receivers). The literature on biosensors is vast, and the first practical demonstration of graphene bioFET-based MC receiver shows promising results in terms of sensitivity, selectivity, and reliability in electrical detection of MC signals \cite{kuscu2020graphene}. However, challenges posed by physiological conditions should be overcome before employing biosensors as electrical interfaces, as detailed in \cite{kuscu2016physical, kuscu2019transmitter}. Conversion from electrical signals into molecular signals is more challenging due to the problem of maintaining continuous molecule generation or supply. Existing electrical stimuli-responsive drug delivery systems rely on limited-capacity reservoirs or polymer chains, e.g., hydrogel, that can store certain types of molecules and release them upon stimulation with a modulated rate. However, these systems are typical irreversible, i.e., they cannot replenish their molecular stock unless they are replaced or reloaded externally \cite{kuscu2019transmitter}.  

In \cite{vanarsdale2020redox}, a redox-based technique is proposed and practically demonstrated for interfacing biological and electronic communication modalities, which can be used to connect a conventional wireless network with engineered bacterial BNTs communicating via molecular signals. The authors introduced the concept of electronically-controlled biological local area network (BioLAN), which includes a biohybrid electrode that transduces information-encoded electronic input signals into biologically-recognized signals in the form of hydrogen peroxide through an oxygen reduction reaction. These signals are recognized by bacterial cells that are attached to the biohybrid electrode, and then biologically propagated across a microbial population with quorum sensing molecules. The overall electronic-biology link is bidirectional such that a microbial subpopulation in the BioLAN generates specific molecules that can be detected by the electrode via an electrochemical oxidation reaction. 

Wearable and epidermal tattoo biosensors and transdermal drug-delivery systems, which have attracted a significant research interest for various healthcare applications, can also be targeted for an macro-nano interface that can connect intrabody IoBNT to the external communication networks, with the integration of communication antennas, such as radio-frequency-identification (RFID)-tag-antennas \cite{kim2018wearable, kiourti2018rfid, wen2020advances, amjadi2018recent}. The challenges lie in the further miniaturization of these devices as well as their continuous operation, since biosensors exposed to physiological fluids suffer contamination, and drug delivery systems require periodic replenishment of their reservoir. 

\subsubsection{\textbf{Optical Interfaces}} 
Light represents an alternative modality to interface the intrabody IoBNT with external networks. In the case that MC is utilized in IoBNT, such an optical interface can be realized with the help of light-sensitive proteins and bioluminescent/fluorescent proteins. 

Optical control of excitable cells, e.g., neurons and muscle cells, can be achieved through a well-known technique called optogenetics \cite{deisseroth2011optogenetics}. The method relies on the genetic modification of natural cells for enabling them to express light-sensitive transmembrane ion channel proteins, e.g., channelrhodopsin. The resulting light-sensitive ion channels open or close depending on the wavelength of the incident photons. The technique enables specificity at the level of single cells in contrast to conventional electrical interfacing techniques, which generally suffer from low level of specificity. It is shown in \cite{hartsough2020optogenetic, liu2018programming} that bacteria can also be engineered to express specific light-sensitive proteins, e.g., bacteriorhodopsin, that pump out protons under illumination, and thus, change the pH of its close environment. 

In \cite{soldner2020survey}, the authors propose that optical control of engineered cells with light-sensitive ion channels can be exploited to enable an optical macro-to-nanoscale interface that can modulate the molecular release of MC transmitters. The authors in \cite{grebenstein2018biological}, experimentally demonstrate that synthetic bacteria expressing bacteriorhodopsin can convert external optical signals to chemical signals in the form of proton concentration at $1$ bit/min conversion rate. They use the same technique to enable an experimental MC testbed in \cite{grebenstein2019molecular}. Similarly, in \cite{balasubramaniam2018wireless}, the authors propose an implantable bio-cyber interface architecture that can enable the \emph{in vivo} optical stimulation of brain cells to control neuronal communications based on external EM signals. Their device architecture includes a wireless antenna unit that connects the implanted device to external networks, an ultrasonic energy harvester, and a micro light emitting diodes ($\mu$-LED) for optical stimulation.

Fluorescent molecules, such as fluorescent proteins, quantum dots, and organic dyes, can also be used to realize a wavelength-selective optical interface. In \cite{kuscu2015fluorescent}, organic dye molecules have been used as nanotransceiver antennas for FRET-based molecular nanonetworks. They act as single molecular optical interfaces that receive optical control signals from an external source and non-radiatively transmit them into a FRET-based nanonetwork. They enable an nano-to-macro interface as well, since the excited fluorescent molecules return to their ground state by releasing a photon at a specific wavelength that can be detected by an external photodetector. Similarly, in \cite{grebenstein2019molecular, nakano2014externally}, it is suggested that a nano-to-macro interface can be realized with engineered bacteria receivers expressing pH-sensitive green fluorescent proteins (GFPs) that change excitation/emission characteristics depending on the pH of the environment. Bioluminescent molecules that are excited upon reaction with a target molecule can also be used for the direct conversion of MC signals to optical signals to enable a nano-to-macro interface, as proposed in \cite{abd2020molcom, kuscu2015internet}.

\subsubsection{\textbf{Other Interfacing Methods}} 
Depending on the communication modality utilized in intrabody IoBNT, there are some other nano-macro interfacing methods proposed in the literature. For example, in \cite{kisseleff2016magnetic}, the authors consider the use of magnetic nanoparticles (MNs) as information carriers in a MC system. They propose a wearable magnetic nanoparticle detector in the form of a ring to connect the intrabody MC to an RF-based backhaul. In a follow-up study \cite{wicke2019magnetic}, they also demonstrate the control of MN-based MC signals in microfluidic channels with external magnetic fields, that could potentially evolve to a bidirectional interface for IoBNT. 

In light of the emerging reports on the EM-based wireless control of cellular functions via specific proteins that are responsive to electromagnetic fields \cite{krishnan2018wireless}, a wireless link is proposed to connect THz-band EM and MC modalities, that can translate into an EM-based nano-macro interface \cite{elayan2020regulating}. The authors in \cite{elayan2020information}, develop an information theoretical model for the mechanotransduction communication channel between an implantable THz nanoantenna acting as the transmitter and a biological protein as the receiver undergoing a conformational change upon stimulation by the THz waves. Although THz-waves are theoretically shown to reliably control the conformational states of proteins, it remains as a challenge to investigate the use of the same modality in sensing the protein states to enable bidirectional wireless interface.  

% \subsection{Big data management and analytics for IoBNT (BDU)}

\subsection{Energy Harvesting, Power Transfer, and Energy Efficiency}
Supply, storage and efficient use of energy is one of the most crucial challenges towards realizing the envisioned IoBNT applications. The energy challenge is currently being addressed through the development of energy harvesting (EH) and wireless power transfer (WPT) techniques to continuously power BNTs, the development of high-capacity energy storage devices at micro/nanoscale, and the design of low-complexity and energy-efficient communication methods for IoBNT. 

For BNTs based on engineered cells, the challenge of energy management is relatively straightforward, as living cells have been evolved over billions of years to make the most efficient use of biochemical energy for realizing vital functionalities. Nonetheless, energy budget requirements of engineered cells may be extended with the introduction of new computation and communication functionalities that are demanded by complex IoBNT applications. On the other hand, there are only a few studies that consider the overall energy requirements of MC for only very simple scenarios \cite{kuran2010energy, bai2014minimum}. The problem is, of course, more challenging for artificial BNTs, such as those that are made up of nanomaterials and missing an inherited metabolism for energy management. 

\subsubsection{\textbf{Energy Harvesting}}
Leaving aside the continuous efforts to reduce the complexity of communication methods for IoBNT, such as modulation and detection techniques \cite{kuscu2019transmitter}, in the hope of increasing energy efficiency, the most promising solution to enable self-sustaining IoBNT is the integration of EH modalities into BNTs. EH has recently received tremendous research interest partly due to the energy requirements set by emerging applications of IoT and IoE. Depending on the application environment and device architectures, various natural energy sources have been considered for harvest by IoT devices \cite{jayakumar2014powering, kuscu2018ioe}. For example, solar energy, vibration sources, electromagnetic sources, e.g., ambient RF EM waves, and metabolic sources have been deemed feasible for harvesting \cite{akan2017internet}. 

Concerning the intrabody and body area applications, human body stands as a vast source of energy in the form of mechanical vibrations resulting from body movements, respiration, heartbeat, and blood flow in vessels, thermal energy resulting from body heat, and biochemical energy resulting from metabolic reactions and physiological processes \cite{dagdeviren2017energy}. Literature now includes a multitude of successful applications of human body EH to power miniature biomedical devices and implants, such as thermoelectric EH from body heat for wearable devices \cite{leonov2013thermoelectric}, vibrational EH from heartbeats \cite{amin2012powering} and respiratory movements \cite{zheng2014vivo} to power pacemakers, as well as biochemical EH from human perspiration \cite{jia2013epidermal}. These together with EH from chemical reactions within the body, such as glucose uptake, lactate release, and pH variations \cite{dagdeviren2017energy, shi2018implantable}, can be exploited to power the BNTs in an intrabody IoBNT. Among the potential EH mechanisms for intrabody IoBNT, mechanical EH has attracted the most interest. Research in this field has gained momentum with the use of flexible piezoelectric nanomaterials, such as ZnO nanowires and lead zirconate titanate (PZT), in nanogenerators, enabling energy harvesting from natural and artificial vibrations with frequencies ranging from very low frequencies ($<1$ Hz) up to several kHz \cite{wang2012nanotechnology, akyildiz2010electromagnetic}. 

\subsubsection{\textbf{Wireless Power Transfer}} 
Another way of powering BNTs and IoBNT applications can be WPT from external sources. WPT has seen significant advances in recent years due to increasing need for powering battery-less IoT devices as well as wearable and implantable devices. Various forms of WPT have been considered for powering medical implants \cite{agarwal2017wireless, ho2014wireless}. For example, near-field resonant inductive coupling (NRIC)-based WPT, the oldest WPT technique, has been in use for widely-used implants, such as cochlear implants \cite{khan2020wireless, manoufali2017near}. Other techniques include near-field capacitive coupling, midfield and far-field EM-based WPT, and acoustic WPT. Power transfer via near-field capacitive and inductive coupling, however, is only efficient for distances on the order of transmitting and receiving device sizes, and for the right alignment of devices, and therefore, might not be suitable for powering micro/nanoscale BNTs \cite{khan2020wireless}. On the other hand, radiative mid-field and far-field EM-based WPTs can have looser restrictions depending on the frequency of EM waves. 

Recent research on mm-wave and THz rectennas suggests the use of high-frequency EM WPT techniques to power BNTs \cite{rong2018nano}. However, for intrabody applications, the higher absorption with increasing frequency and power restrictions should be taken into account. Nonetheless, simultaneous wireless information and power transfer techniques (SWIPT) utilizing THz-band have been investigated for EM nanonetworks \cite{feng2020dynamic, rong2017simultaneous}. Similar SWIPT applications have been considered for MC, where the receiving BNTs use the received molecules for both decoding the information and energy harvesting \cite{guo2017smiet, deng2016enabling}. There are also applications of acoustic WPT for biomedical implants using external ultrasonic devices \cite{maleki2011ultrasonically, larson2011miniature}. Although not implemented yet, ultrasonic EH has been also considered for powering BNTs with piezoelectric transducers \cite{donohoe2016powering, donohoe2017nanodevice, balasubramaniam2018wireless}.

An interesting research direction in parallel with the wider IoE vision is towards hybrid EH systems that can exploit multiple energy sources. Prototypes have been implemented for ZnO nanowire-based hybrid cells for concurrent harvesting of solar and mechanical energies \cite{xu2009nanowire}, and piezoelectric PVDF-nanofiber NG based hybrid cells for biomechanical and biochemical EH from bodily fluids \cite{hansen2010hybrid}. A hybrid EH architecture is also proposed for IoE comprising modules for EH from light, mechanical, thermal, and EM sources \cite{akan2018internet}. The same hybrid architectures could be considered for IoBNT as well to maintain the continuous operation of BNTs. EH from multiple resources can reduce the variance at power output with the addition of alternative complementary procedures in a modular fashion as investigated in \cite{akan2018internet}.

\subsubsection{\textbf{Energy Storage}}
Storage of energy is also important when BNTs cannot continuously satisfy their power requirements via EH and WPT techniques. There has been considerable interest in miniaturizing the energy storage technologies to make them size-compatible with MEMS and NEMS devices. Some of the efforts have been devoted to develop micro-batteries, miniaturized versions of conventional thin-film lithium-ion batteries, benefiting from novel nanomaterials \cite{albano2008fully}. There are even a few studies focusing on nanoscale versions of lithium batteries \cite{vullum2006characterization, akyildiz2010electromagnetic}. However, they suffer from low energy density, short lifetimes, and potential toxicity in \emph{in vivo} applications. More promising solution is micro-supercapacitors (MSCs), which provide significantly higher energy storage capacity, higher charge/discharge rates, and more importantly, scalability and flexibility, which are crucial for their integration into BNTs \cite{zhang2019recent, patil2019status, lu2019wearable}. 

Several types of materials have been considered for the design of MSC electrode to improve the energy density. Carbon nanomaterials, such as CNTs and graphene, are the most widely researched materials due to their abundance and stability, which is reflected to an extended lifetime \cite{zhang2019recent}. Due to its extremely high surface-to-volume ratio, high mobility and flexibility, graphene has attracted particular attention \cite{liu2010graphene, wu2014recent}. Additionally, conducting polymers, such as PEDOT/PSS, and graphene/conducting polymer heterostructures are also considered as flexible electrodes for MSCs \cite{zhang2019recent}. The research in MSCs is still at early stages; however, we believe that with the increase of energy density and further reduction of sizes, they can be a viable candidate for energy storage units in BNTs.

% \subsection{Security and privacy aspects of IoBNT (Confidentiality, integrity, availability, authentication) (BDU)}

% \cite{loscri2014security}

% \subsubsection{Artificial immune systems}

% \subsubsection{Swarm security}

% \subsubsection{DNA-inspired encryption techniques}

\subsection{Biocompatibility and Co-existence}
Biological processes are complex, and intertwined, often through intricate relationships that are yet to be uncovered. Perturbation of homeostasis maintained by these relationships may result in serious disorders. Even more complicated is the fact that the composition of physiological environment and the interactome may have a large variance among different members of the same species. For example, gut microbiome is known to be composed of different types of bacteria in different people \cite{suzuki2020role}. Therefore, the evaluation of \emph{in vivo} IoBNT applications in terms of biocompatibility is very challenging, however, must be considered seriously.

Biocompatibility constraints for IoBNT can be viewed from two angles \cite{egan2018coexistence}. First, an IoBNT application, along with all the communication methods and devices therein, should not disrupt the homeostasis of the organism it is implemented in. Such disruption might occur when the introduced application has toxic, injurious, or adverse effects on the living cells and biochemical processes. Second, an implanted IoBNT application should be able to operate without its performance being degraded by the co-existing biochemical processes. Performance degradation usually follows when an IoBNT application alters the metabolic activities, because such alternation invokes the immune response that might in turn lead to the  \emph{rejection} of the deployed application. Rejection can occur in the form of expulsion of the IoBNT application from the organism, encapsulation of the BNTs with biological cells and tissues, or inflammation or death of the surrounding tissues. In the case of MC, performance degradation may also happen as a result of cross-talk caused by the natural biochemical signaling.

Biocompatibility concerns both materials used in the physical architecture of BNTs, and the networking, energy harvesting, power transfer, and interfacing processes of the IoBNT. In terms of materials, synthetic biology-based BNTs can be considered highly biocompatible, as they adopt living cells and cellular components as the substrate \cite{unluturk2015genetically}. Likewise, fluorescent protein- and DNA-based BNTs are also of biological origins, and thus, can be expected to offer similar levels of biocompatibility \cite{kuscu2015internet}. However, depending on their exact biological origin and their overall amount in the body, they may still trigger the immune response. For example, a virus-based synthetic BNT can be labeled as foreign agent and attacked by the immune system, unless it is designed to possess a kind of stealth proteins that help escape the immune control \cite{schmid2018adenoviral}. For artificial BNTs based on nanomaterials, biocompatibility is more challenging. There is still no consensus on a universal test of biocompatibility for nanomaterials, leading to conflicting results in the literature about almost all materials. Complexity of the biological systems and reproducibility problem for \emph{in vivo} and \emph{in vitro} tests are the main causes of the ongoing ambiguity. Nonetheless, many polymers (e.g., PMMA, Parylene), gold, titanium, and some ceramics are widely known to be biocompatible \cite{chen2008biocompatible, kulkarni2015titanium, treccani2013functionalized}. Carbon-based materials, e.g., CNT and graphene, have been reported as both biocompatible and toxic in different works, preventing a generalization over these nanomaterials. This is attributed to large variations in their physicochemical properties, e.g., size, shape, surface characteristics, adopted in different works \cite{bianco2013graphene, ferrari2015science}. However, it has been repeatedly reported that their biocompatibility can be modulated with chemical manipulation \cite{bianco2013graphene}. For example, surface functionalization with dextran is shown to reduce the toxicity of graphene oxide (GO), hinting at strategies to make carbon-based nanomaterials suitable for safe \emph{in vivo} applications \cite{zhang2011vitro, ferrari2015science}. Similarly, nanoparticles are shown to be detoxified upon the functionalization of their surfaces with smart/benign ligands \cite{alkilany2010toxicity}. 

In terms of processes, attention must be paid to the communication, bio-cyber interfacing, energy harvesting, transfer and storage processes.  In EM-based and acoustic communication and power transfer processes, for example, the biocompatibility is crucial for preventing the damage on biochemical structures, e.g., tissue damage by heating, and closely linked to the frequency and power of the EM or acoustic waves. For human body applications, the exposure limits are regulated by the Food and Drug Administration (FDA) in US, as 100 $\mu$W/mm$^2$ for RF waves and 7.2 mW/mm$^2$ for ultrasound waves, \cite{balasubramaniam2018wireless, okwori2018micro}. These limits should be taken into account in the design of IoBNT technologies.  

In the cases that MC is adopted in IoBNT applications, the concentration and type of molecules used for communications is very critical for biocompatibility. First, the information-carrying molecules should not invoke the immune response. The degradation of information molecules by enzymes should also be taken into account to prevent performance degradation. Also, the MC signals should not interfere with the inherent biological systems. The type of information molecules must be orthogonal to the molecules involved in biological processes to prevent interference, or their concentration should be low enough not to disrupt these processes. This so-called co-existence challenge has recently attracted close attention of MC researchers, who suggest different solution strategies \cite{akdeniz2019reactive, egan2019estimation, egan2018strategies}. For example, in \cite{egan2018strategies}, a cognitive radio-inspired transmission control scheme is proposed to overcome interference between MC networks and co-existing biological networks. In \cite{kuscu2019channel}, various channel sensing methods inspired from the spectrum sensing techniques in EM cognitive radio are proposed to estimate the instantaneous composition of the MC channel with ligand receptors in terms of molecule types. The effect of biological cross-talk on the MC is also investigated in \cite{kuscu2020detection}, where the performance of several detection methods are investigated in terms of their ability to ensure reliability under such biological interference. 

\section{Conclusion}
In this survey, a comprehensive overview of IoBNT framework along with its main components and applications is provided to contribute to an holistic understanding of the current technological challenges and potential research directions in this emerging field. In light of rapid advances in synthetic biology, nanotechnology, and non-conventional communications made possible by interdisciplinary approaches, we believe that the enormous potential of the IoBNT will soon be realized with high-impact medical, environmental and industrial applications. 

\bibliographystyle{ieeetran}
\bibliography{IoBNT_new}

\end{document}